\documentclass[11pt,a4paper]{article}
\pdfoutput=1
\usepackage{jheppub} 
\usepackage{xspace}
\usepackage{subfigure}
\usepackage{amsmath}
\usepackage{amssymb}
\usepackage{amsfonts}
\usepackage{mathrsfs}

\newcommand{\be}{\begin{equation}}
\newcommand{\ee}{\end{equation}}

\newcommand{\bea}{\begin{eqnarray}}
\newcommand{\eea}{\end{eqnarray}}
\newcommand{\beanon}{\begin{eqnarray*}}
\newcommand{\eeanon}{\end{eqnarray*}}
\newcommand{\ba}{\begin{array}}
\newcommand{\ea}{\end{array}}
\newcommand{\bd}{\begin{description}}
\newcommand{\ed}{\end{description}}
\newcommand{\bi}{\begin{itemize}}
\newcommand{\ei}{\end{itemize}}
\newcommand{\ben}{\begin{enumerate}}
\newcommand{\een}{\end{enumerate}}
\newcommand{\bc}{\begin{center}}
\newcommand{\ec}{\end{center}}

\newcommand{\eqn}[1]{eq.(\ref{#1})}

\newcommand{\tbn}[1]{tab.~\ref{#1}}

\newcommand{\fig}[1]{fig.~\ref{#1}}
\newcommand{\figs}[2]{figs.~\ref{#1}--\ref{#2}}
\newcommand{\figsc}[2]{figs.~\ref{#1},~\ref{#2}}
\newcommand{\sect}[1]{sect.~\ref{#1}}

\newcommand{\rf}[1]{ref.~\cite{#1}}
\newcommand{\al}{\alpha}

\newcommand{\hzero}{\ensuremath{h}} %light neutral Higgs
\newcommand{\Hzero}{\ensuremath{H}} %light neutral Higgs

\title{
Interference effects in Heavy Higgs production via gluon fusion in the Singlet Extension of the
Standard Model.
}

\author[a,b]{Ezio Maina}

\affiliation[a]{INFN, Sezione di Torino,\\
Via Giuria 1, 10125 Torino, Italy}
\affiliation[b]{Dipartimento di Fisica, Universit\`a di Torino,\\
Via Giuria 1, 10125 Torino, Italy}

\emailAdd{maina@to.infn.it}

\abstract{
The measurements of the properties of the Higgs boson still leave room for a non minimal scalar sector.  
Extensions of the Standard Model typically involve multiple neutral Higgs fields which can interfere among
themselves. We show that these interference effects can be substantial taking as example the one Higgs Singlet Model,  the simplest renormalizable addition to the SM.
}

\begin{document}

\maketitle
        
\section{Introduction}
\label{sec:intro}
All mesurements performed in LHC Run I relative to the resonance 
discovered at about 125 GeV 
\cite{Aad:2012tfa,Chatrchyan:2012ufa}
are consistent with the hypothesis that
the new particle is indeed the Standard Model Higgs boson.
While the mass is already
known to an astonishing three per mill from the latest
published CMS measurement \cite{Khachatryan:2014jba}, the signal strengths
$\mu^i = \sigma^i/\sigma^i_{SM}$, where $i$ runs over the decay channnels, 
are known to about 10 to 20\% \cite{Khachatryan:2014jba,ATLAS:2014yyy,ATLAS:2014bny}.
This leaves room for modifications of the SM with a more complicated Higgs sector
provided they are consistent within experimental errors with the data. 
A larger Higgs sector implies that additional scalar states are present in the spectrum.
Direct searches have provided limits on the existence of new spin zero particles
and on the strengths of their couplings \cite{CMS:2014www}.
The larger luminosity and energy in Run II will allow more precise measurements of the
already discovered
Higgs properties and extend the mass range in which other scalars can be searched for.

The simplest renormalizable extension
of the SM is the one Higgs Singlet Model (1HSM). It introduces one 
additional real scalar field which is a singlet under all SM gauge groups. 
The 1HSM has been extensively investigated in the literature 
\cite{Silveira:1985rk,Schabinger:2005ei,O'Connell:2006wi,BahatTreidel:2006kx,
Barger:2007im,Bhattacharyya:2007pb,Gonderinger:2009jp,Dawson:2009yx,Bock:2010nz,
Fox:2011qc,Englert:2011yb,Englert:2011us,Batell:2011pz,Englert:2011aa,Gupta:2011gd,
Batell:2012mj,Pruna:2013bma,Lopez-Val:2014jva,Englert:2014ffa,Chen:2014ask,Robens:2015gla,
Logan:2014ppa},
however, to our knowledge,
no public MC implementation of the model is available.
In this note we present such an implementation using FeynRules for the derivation of the vertices
and Madgraph 5 for the generation of the amplitudes. 
We then discuss the simple case of Higgs production via gluon fusion at the LHC.

\section{The Singlet Extension of the Standard Model}
\label{sec:model}
The singlet extension of the SM is defined by adding to the standard Lagrangian the following 
gauge invariant, renormalizable term:
\begin{equation}\label{lag:s}
\mathscr{L}_s = 
\partial^{\mu} S \partial_{\mu} S 
 -\mu_1^2 \Phi^{\dagger} \Phi -\mu_2^2 S ^2 + \lambda_1
(\Phi^{\dagger} \Phi)^2 + \lambda_2  S^4 + \lambda_3 \Phi^{\dagger}
\Phi S ^2.
\end{equation}
where  $S$ is a real {$SU(2)_L\otimes\,U(1)_Y$} singlet  and
$\Phi$, is the SM Higgs {weak isospin} doublet. Here and in the following we adopt the notation
of \rf{Pruna:2013bma}.
A $\mathcal{Z}_2$ symmetry 
which forbids additional terms in the potential is assumed. A detailed discussion of the 1HSM without
$\mathcal{Z}_2$ symmetry can be found in \rf{Chen:2014ask}.

The neutral components of these fields can be expanded around their
respective {Vacuum Expectation Values (VEVs)} as follows:

\begin{equation}
 \Phi = \left(\begin{array}{c}  
  G^{\pm} \\ \cfrac{v_d + l^0 + iG^0}{\sqrt{2}}
 \end{array}\right) \qquad \qquad S = \cfrac{v_s + s^0}{\sqrt{2}}
 \label{eq:components}.
\end{equation}

\noindent The minimum of the potential is achieved for 

\begin{alignat}{5}
 \mu^2_1 = \lambda_1 v_d^2 + \cfrac{\lambda_3 v_s^2}{2}; \qquad \qquad 
  \mu^2_2 = \lambda_2 v_s^2 + \cfrac{\lambda_3 v^2_d}{2}
  \label{eq:minimum},
\end{alignat}
provided
\begin{alignat}{5}
 \lambda_1, \lambda_2 > 0; \qquad 4\lambda_1\lambda_2 - \lambda_3^2 > 0 \label{eq:stability}\; .
\end{alignat}

The  mass matrix in the gauge basis 
can be diagonalized into the (tree--level) mass basis introducing new fields:
\begin{alignat}{5}
\hzero =l^0  \cos\alpha  -s^0 \sin\alpha  \qquad \text{and} \qquad
 \Hzero &= l^0 \sin\alpha  +s^0 \cos\alpha 
\label{eq:masseigen-repeat}
\end{alignat}
with $-\frac{\pi}{2} < \alpha < \frac{\pi}{2}$.

\noindent The  masses are

\be
 m^2_{\hzero,\Hzero} = \lambda_1\,v_d^2 + 
 \lambda_2\,v_s^2 \mp |\lambda_1\,v_d^2 - 
 \lambda_2\,v_s^2|\,\sqrt{1+\tan^2(2\alpha)}\, ,
\quad 
 \tan(2\alpha) = \cfrac{\lambda_3v_dv_s}{\lambda_1 v_d^2 - \lambda_2v_s^2}\, ,
 \label{eq:masseigen}
\ee
\noindent  with the convention $m_{\Hzero}^2 > m_{\hzero}^2$.
They correspond to a 
light [$\hzero$] and a heavy [$\Hzero$] $\mathcal{CP}$-even  mass--eigenstate. 

The Higgs sector in this model is determined by five independent parameters,  which can be chosen as
\be
\label{eq:parameters}
m_{\hzero},\,m_{\Hzero}, \,\sin\al, \,{v_d}, \,\tan\beta \,\equiv\,{v_d}/{v_s}\, ,
\ee 
where the doublet VEV is fixed in terms of the Fermi constant through $v_d^2 = G_F^{-1}/\sqrt{2}$. 
Furthermore one of the Higgs masses is determined by the LHC measurement of $125.02$ GeV.
Therefore, three parameters of the model are presently free. 

\begin{figure}[tb]
\centering
\subfigure{	 
\hspace*{-2.3cm} 
\includegraphics*[width=8.3cm]{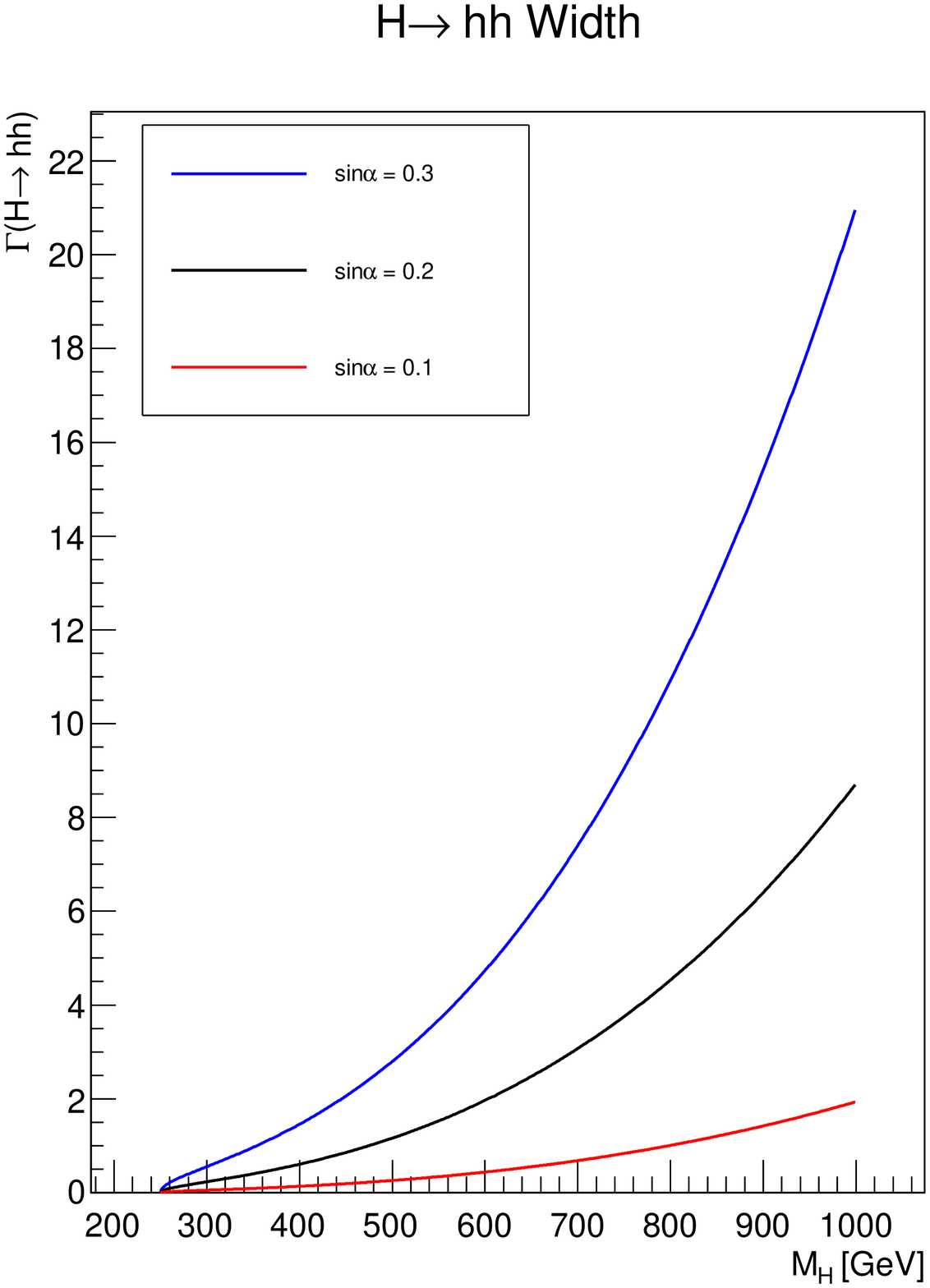}
%\hspace*{-0.7cm}
\includegraphics*[width=8.3cm]{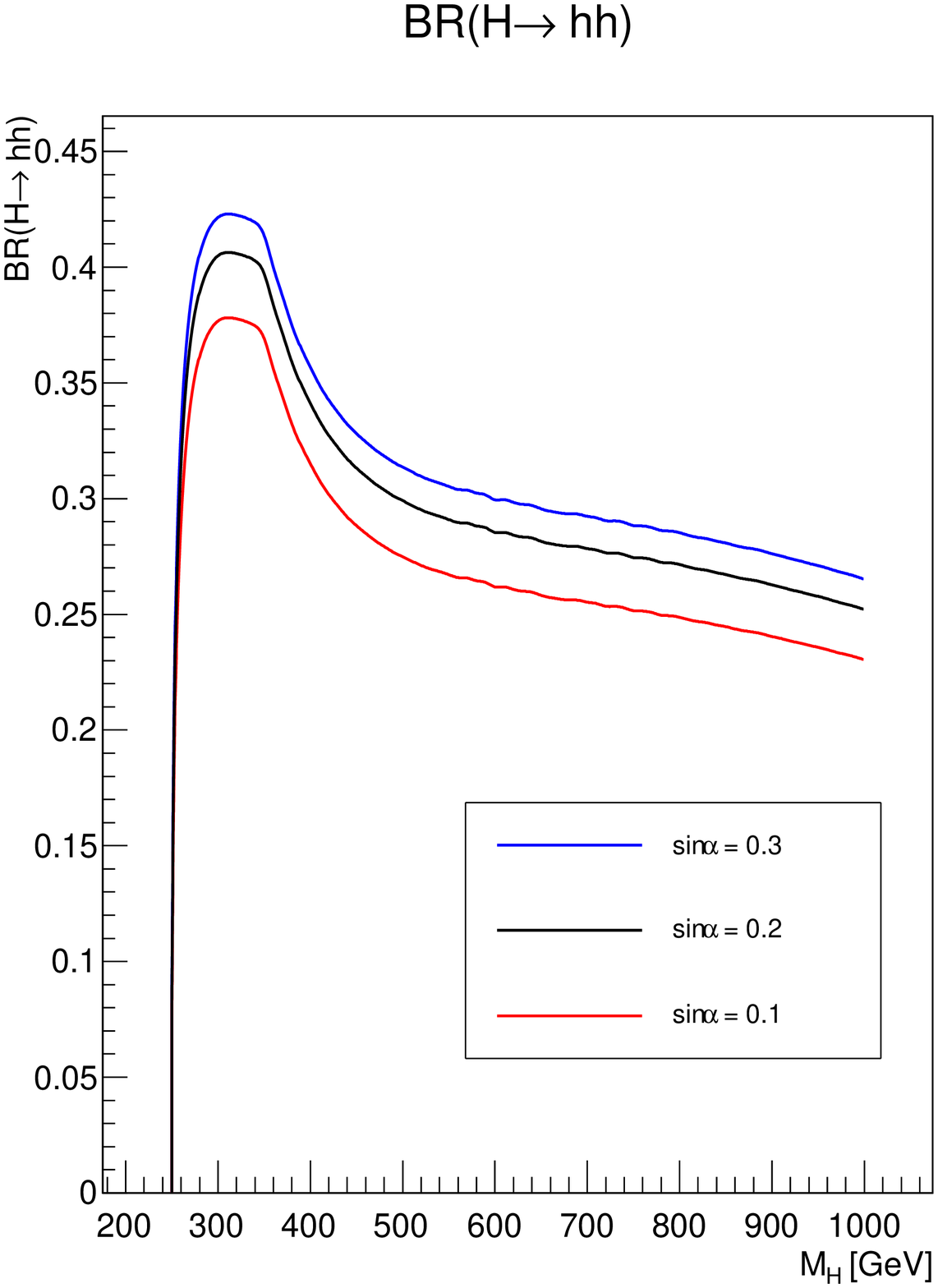}
\hspace*{-3cm}
}
\caption{On the left: the partial width for the $H \rightarrow hh$ process for $s_\alpha = 0.3,\, 0.2,\,
 0.1$ and $\tan\beta = 1.0$.
On the right: the corresponding branching ratios.
}
\label{fig:BRH2hh}
\end{figure}

As only the doublet component, before mixing,
can couple to the fermions (via ordinary Yukawa interactions) and the gauge bosons
(via the gauge covariant derivative), all of the corresponding Higgs couplings
are rescaled universally, yielding

\begin{alignat}{5}
 g_{xxs} = g_{xxh}^{\text{SM}}(1 + \Delta_{xs}) \qquad \text{with} \qquad 1+\Delta_{xs} = 
 \begin{cases}\cos\alpha& s\,=\,\hzero\\ \sin\al&s=\Hzero \end{cases} \label{eq:coupling}. 
\end{alignat}

\begin{alignat}{5}
 g_{xxs_1s_2} = g_{xxhh}^{\text{SM}}(1 + \Delta_{xs_1}) (1 + \Delta_{xs_2})  \label{eq:coupling2}. 
\end{alignat}

where $xx$ represents a pair of SM fermions or vectors.

The couplings in the scalar sector involve $\tan\beta$ and are slightly more complicated. As an example
we reproduce the triple scalar vertices in terms of the independent parameters mentioned above
($s_{\alpha}=\sin\alpha$, $c_{\alpha}=\cos\alpha$):

\begin{align}
\label{eq:3hvertices}
V_{hhh}
 & \qquad   -\frac{3 i e M_h^2} {2 M_{W} s_{W}}  \{ c_{\alpha}^3 - s_{\alpha}^3 \tan \beta \}    \\
V_{hHH}
 & \qquad  -\frac{ i e s_{\alpha} c_{\alpha}} {2 M_{W} s_{W}} \{ s_{\alpha}-c_{\alpha}\tan\beta \} \{ M_h^2 +
  2 M_H^2 \}      \\
V_{Hhh} & \qquad  -\frac{ i e s_{\alpha} c_{\alpha}} {2 M_{W} s_{W}}  \{c_{\alpha} + s_{\alpha} \tan\beta\} \}
 \{ 2 M_h^2 + M_H^2 \}     \\ 
V_{HHH} &  \qquad -\frac{3 i e M_H^2}{2 M_{W} s_{W}} \{ s_{\alpha}^3 + c_{\alpha}^3 \tan\beta \} 
\end{align} 

The tree level partial width for the decay of the heavy scalar into two light ones reads:
\begin{equation}
\label{eq:H2hhWidth}
\Gamma (H\rightarrow h h ) = \frac{e^2 M_H^3}{128 \pi M_W^2 s_W^2}
\left( 1 - \frac{4M_h^2}{M_H^2} \right)^{\frac{1}{2}}
\left( 1 + \frac{2M_h^2}{M_H^2} \right)^2
s_{\alpha}^2 c_{\alpha}^2 \left( c_{\alpha} + s_{\alpha} \tan\beta \right) ^2
\end{equation}

In \fig{fig:BRH2hh} we show the partial width and the corresponding branching ratios as a function of the
heavy Higgs mass for 
$s_\alpha = 0.3,\, 0.2,\, 0.1$ and $\tan\beta = 1.0$. The BR is computed as the ratio of the lowest order 
width in \eqn{eq:H2hhWidth} to the total width given in \rf{Heinemeyer:2013tqa} multiplied by
$s^2_\alpha$.
The BR rises sharply above the kinematical threshold and, for the parameter range we have considered,
remains larger than 25\% up to $M_H = 1$ TeV. This raises the tantalizing prospect of a relatively
abundant production of heavy Higgses followed by their decay into two light ones if $M_H > 2 M_h$.
See \rf{Chen:2014ask} for a detailed discussion in the framework of the 1SHM.

\section{Limits on the parameters}
\label{sec:limits}
The strongest limits on the parameters of the 1HSM come from measurements of the coupling strengths
of the light Higgs 
\cite{Khachatryan:2014jba,ATLAS:2014yyy,ATLAS:2014bny,CMS:2014www},
which dominate
for small masses of the heavy Higgs,
and from the contribution of higher order corrections to precision measurements, in
particular to the mass of the W boson \cite{Lopez-Val:2014jva}, which provides the tightest constraint
for large $M_H$.
The most precise result for the overall coupling strength of the Higgs boson from CMS
\cite{Khachatryan:2014jba} reads

\be
\hat{\mu} = \hat{\sigma}/\sigma_{SM} = 1.00 \pm 0.13.
\ee
Therefore the absolute value of $\sin\alpha$ cannot be larger than about 0.4. This is in agreement with
the limits obtained in \rf{Lopez-Val:2014jva,Robens:2015gla} which conclude that the largest possible
value for the absolute value of $\sin\alpha$ is 0.46 for $M_H$ between 160 and 180 GeV.  This limit
becomes slowly more stringent for increasing heavy Higgs masses reaching about 0.2 at $M_H = 700$
GeV.

\section{Interference effect and simulation tools}
\label{sec:interference}

The focal point of this note is the interference between the two Higgs fields. In general, any amplitude involving a single Higgs exchange  can be written as
\begin{equation}
\label{eq:higgsExchange}
A = A^\prime
 \left( 
\frac{c_{\alpha}^2}{q^2-M_h^2 +i \Gamma_h M_h}
+
\frac{s_{\alpha}^2}{q^2-M_H^2 +i \Gamma_H M_H}
\right)       + A_0 = A_1 + A_0
\end{equation}
where $A_0$ does not involve the scalar fields.
The real parts of the two propagators intefere destructively for $M_h^2 < q^2 < M_H^2$
and constructively for $q^2 < M_h^2$ and $M_H^2 < q^2$. This phenomenon has already been noticed in \rf{Englert:2014ffa} where however it was dismissed as numerically irrelevant.

As will be argued in \sect{sec:results}, the interference effects can be substantial, their relevance
increasing with $M_H$.
The shape of the the heavier Higgs peak is also strongly affected and this will need to be taken into
account in any search for additional scalars and eventually in the measurement of their properties.

These features are neglected by any prediction based
on the narrow width approximation, or equivalently on a production times decay approach.

The relevance of the interference term does depend on the relative size of $A_0$ and $A_1$ in
\eqn{eq:higgsExchange} and is expected to be significant for processes in which the resonant part
of the amplitude is large as in gluon fusion for large values of the heavy Higgs mass.

Clearly, this interference between different Higgs fields is not a peculiarity of the Singlet Model. It will
indeed occur in any theory with  multiple scalars which couple to the same set of elementary particles, albeit possibly with different strengths.

In order to allow for the Monte Carlo simulation of the 1HSM we have used FeynRules 
\cite{Christensen:2008py,Alloul:2013bka} to prepare a UFO file \cite{Degrande:2011ua} for the model,
which can be imported, as we did,  in MadGraph 5 \cite{Alwall:2014hca} and many other general purpose
MC tools. It enables the simulation at tree level of any process in the 1HSM. The UFO file can be downloaded from http://personalpages.to.infn.it/$\scriptstyle \sim$maina/Singlet.

The gluon fusion channel in MadGraph requires particular care. MadGraph treats the 
gluon-Higgs effective vertex
in the narrow width approximation, through an expansion of the top loop amplitude in powers
of $M_h^2/M^2_{top}$, which is unsuitable in the present context and which fails for Higgs masses above
the t-tbar threshold. The appropriate effective vertex $V_{ggh}$
must be introduced by hand in the matrix element:
\be
\label{eq:ggheff}
V_{ggh} = -i \frac {2\, \alpha_s (\sqrt{2}\, G_F)^{\frac{1}{2}}  M_{top}^2} {\pi \hat s}
                   (1-\frac{1}{2}(1-\tau) \,C(\tau)),
\ee
with                   
\begin{alignat}{5}
          C(\tau) = \begin{cases} -2\,\arcsin \left( 1/\sqrt{\tau}\right)^2 & \tau > 1\\
                         \frac{1}{2} \left(  \log\left(\frac{1+\beta}{1-\beta}\right) -i \pi\right )^2
                         & \tau < 1
                         \end{cases}
\end{alignat}
where $\tau = 4 M_{top}^2/\hat s$, $\beta = \sqrt{1-\tau}$ and $\hat{s}$ is the square of the sum of the momenta carried by the gluons which in general is not equal to the Higgs mass squared.

The tensor structure of the vertex is already taken care of by MadGraph.

\section{Results}
\label{sec:results}

As an example we have studied Higgs production through gluon fusion in the four electron channel,
$gg \rightarrow h,H \rightarrow ZZ \rightarrow 4e$, at the LHC with a center of mass energy of 13 TeV.

We neglect the non resonant contribution given by $q \bar{q} \rightarrow Z Z$ and by 
$g g \rightarrow Z Z$ through a quark box
amplitude. This continuum is known to be large and in particular there is a non negligible
interference between
the box contribution and Higgs production through the heavy quark three point loop.
These terms are well known
\cite{Glover:1988fe,Glover:1988rg,Binoth:2006mf,Accomando:2007xc,Campbell:2011cu,Kauer:2012ma,
Dixon:2003yb,Dixon:2008xc,Accomando:2011eu,Campbell:2011bn,Frederix:2011ss,Melia:2012zg,
Agrawal:2012df,Kauer:2012hd} and are essential for accurate phenomenological predictions.
The region of large invariant masses of  the four final state leptons in  
$gg \rightarrow ZZ,WW \rightarrow 4l$ has been studied in detail in \rf{Kauer:2012hd} in the SM with a
Higgs of 125 GeV. 
Above the light Higgs peak,
the differential cross section is dominated by the continuum for $M_{4l} < 2 M_{top}$.
The box contribution however drops more rapidly with increasing  $M_{4l}$ than the Higgs mediated one.
Unitarity requires the interference between these two components to be negative and while it is essentially 
negligible below the top threshold it becomes more relevant for larger masses and exceeds 50\% of the incoherent sum of the two contributions in the one TeV range.

\begin{figure}[tbh]
\includegraphics*[width=12.3cm]{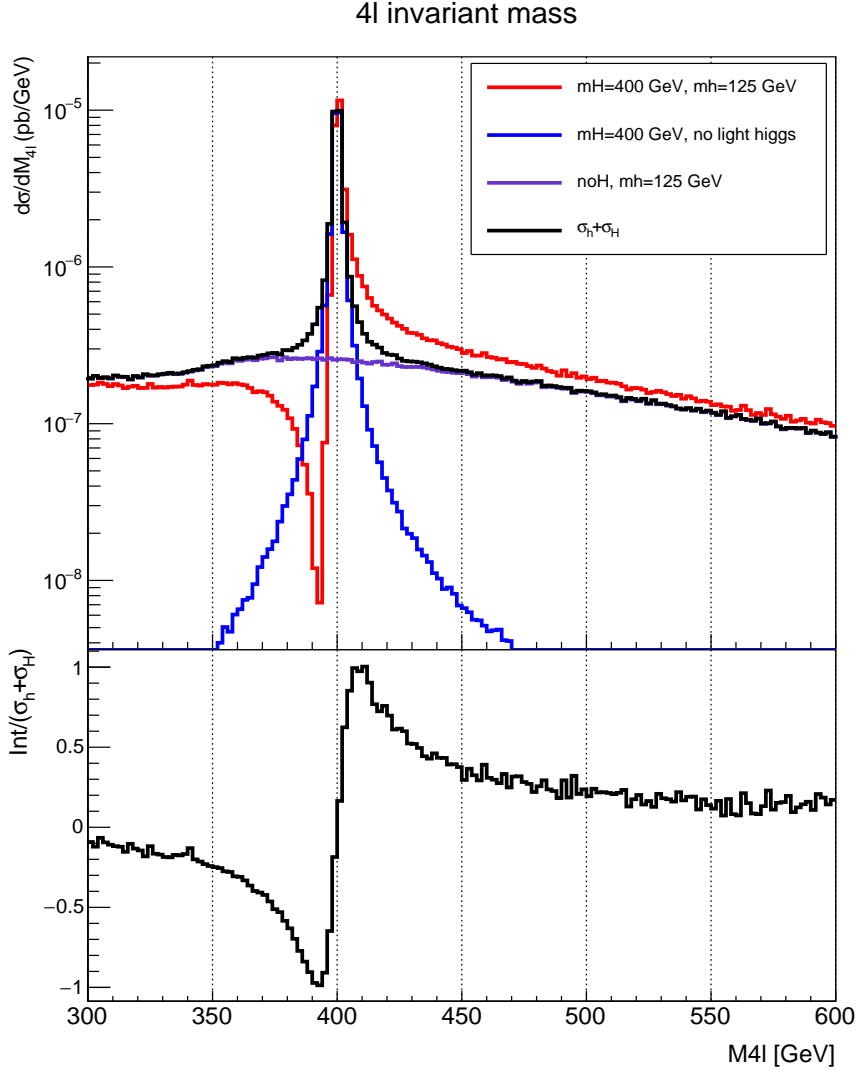}
\caption{ $gg \rightarrow h,H \rightarrow ZZ \rightarrow 4e$
at the LHC with a center of mass energy of 13 TeV.
Upper part:four lepton invariant mass distribution for $M_h=125$ GeV, $M_H=400$ GeV and
 $s_\alpha= 0.2$. The red line is the full result ($\sigma$). 
The violet histograms ($\sigma_h$) shows the SM prediction with
Higgs couplings scaled by $\cos\alpha$. The blue line ($\sigma_H$) 
gives the result when the light Higgs diagrams
are neglected while the $H\rightarrow h h$ contribution to $\Gamma_H$ is retained. The black histogram
($\sigma_{NI}$) refers to the sum of the violet and blue lines and corresponds to neglecting the
 interference effects.
Lower part: the ratio $\frac{\sigma - \sigma_{NI}}{\sigma_{NI}}$.
}
\label{fig:4lmass_400_02}
\end{figure}
\vspace{5mm}

\begin{figure}[tbh]
\includegraphics*[width=12.3cm]{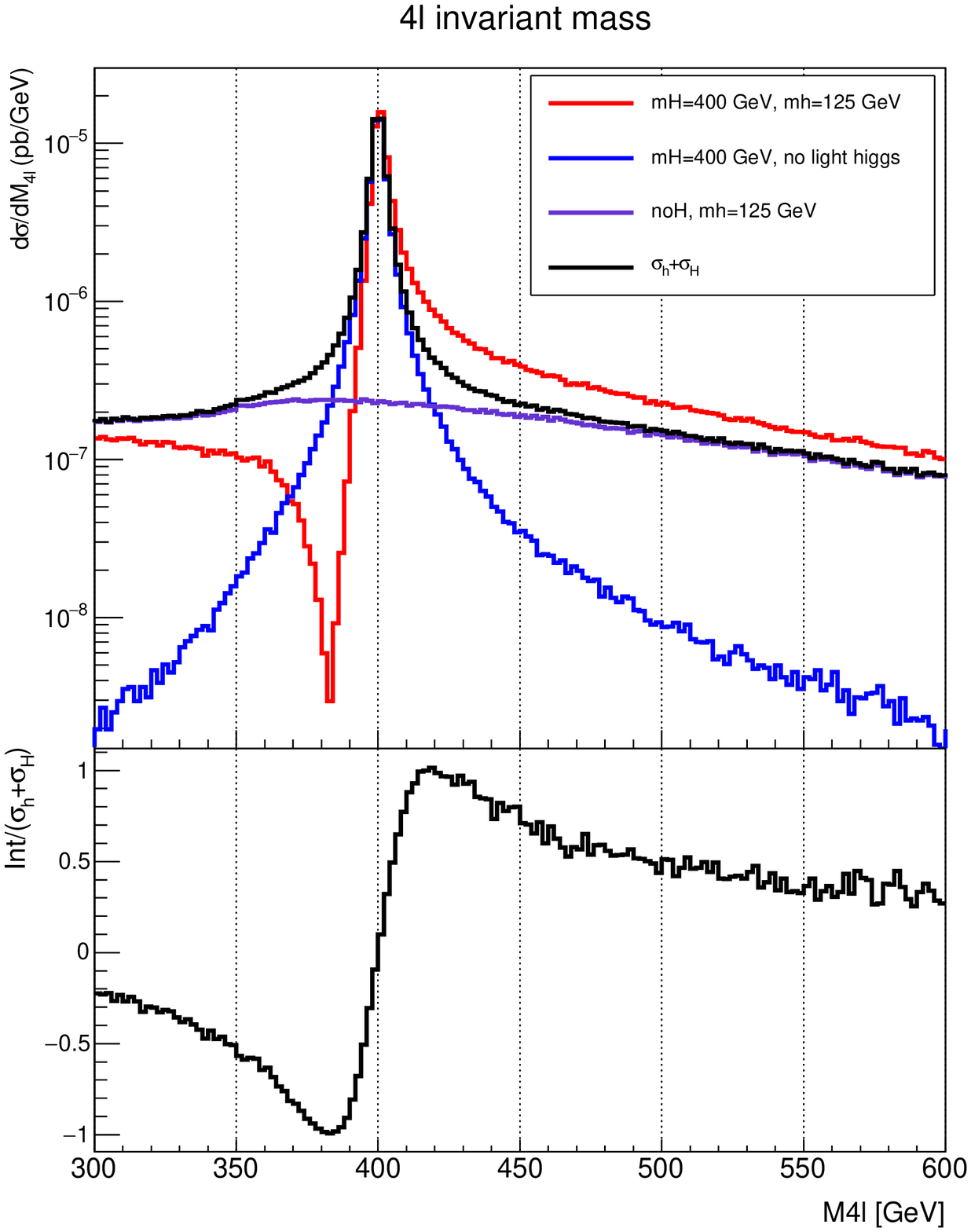}
\caption{Four lepton invariant mass distribution for $M_h=125$ GeV, $M_H=400$ GeV, $s_\alpha= 0.3$.
The meaning of the various histograms is as in \fig{fig:4lmass_400_02}.
}
\label{fig:4lmass_400_03}
\end{figure}

\begin{figure}[tbh]
\includegraphics*[width=12.3cm]{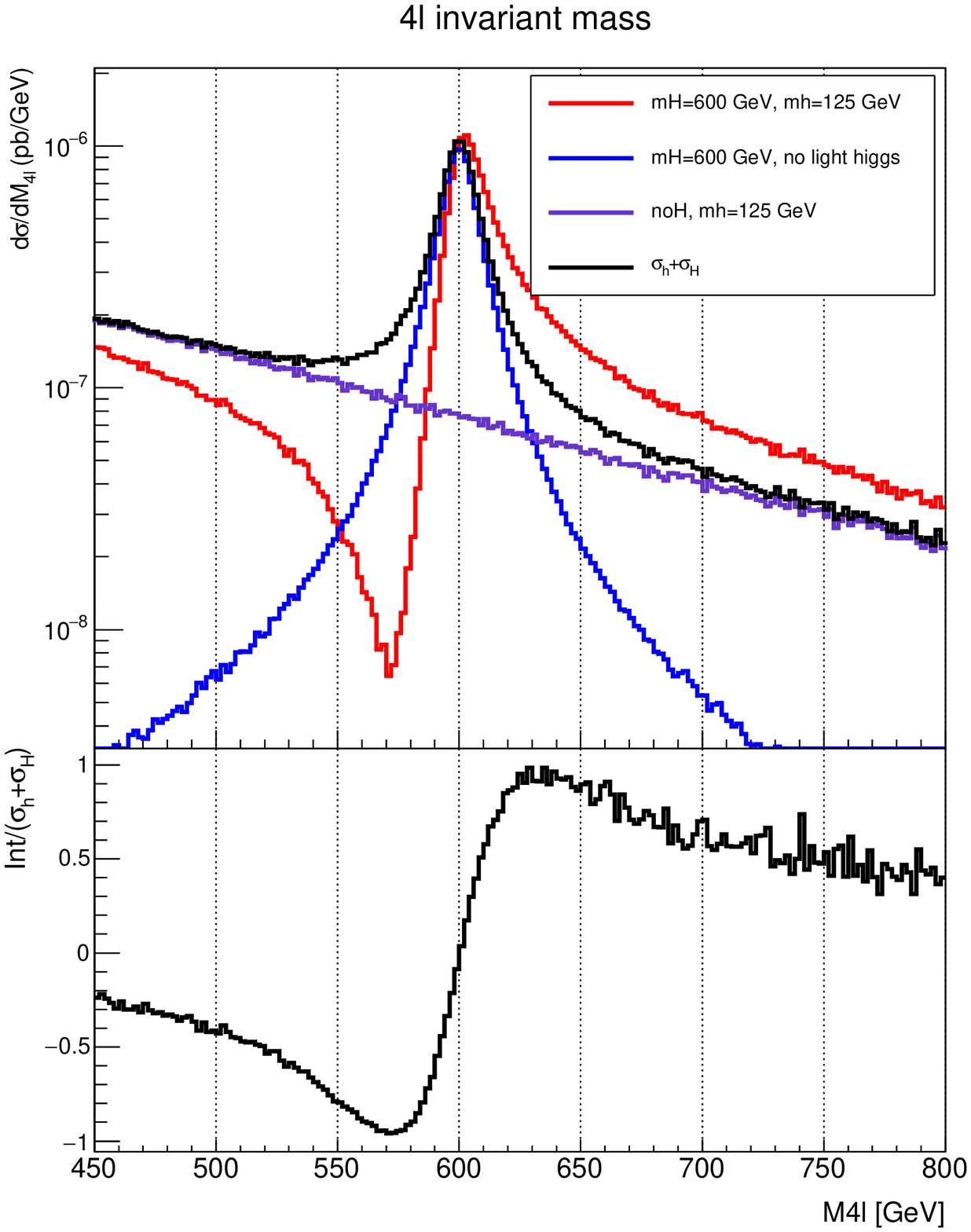}
\caption{Four lepton invariant mass distribution for $M_h=125$ GeV, $M_H=600$ GeV, $s_\alpha= 0.3$.
The meaning of the various histograms is as in \fig{fig:4lmass_400_02}.
}
\label{fig:4lmass_600_03}
\end{figure}

\begin{figure}[tbh]
\includegraphics*[width=12.3cm]{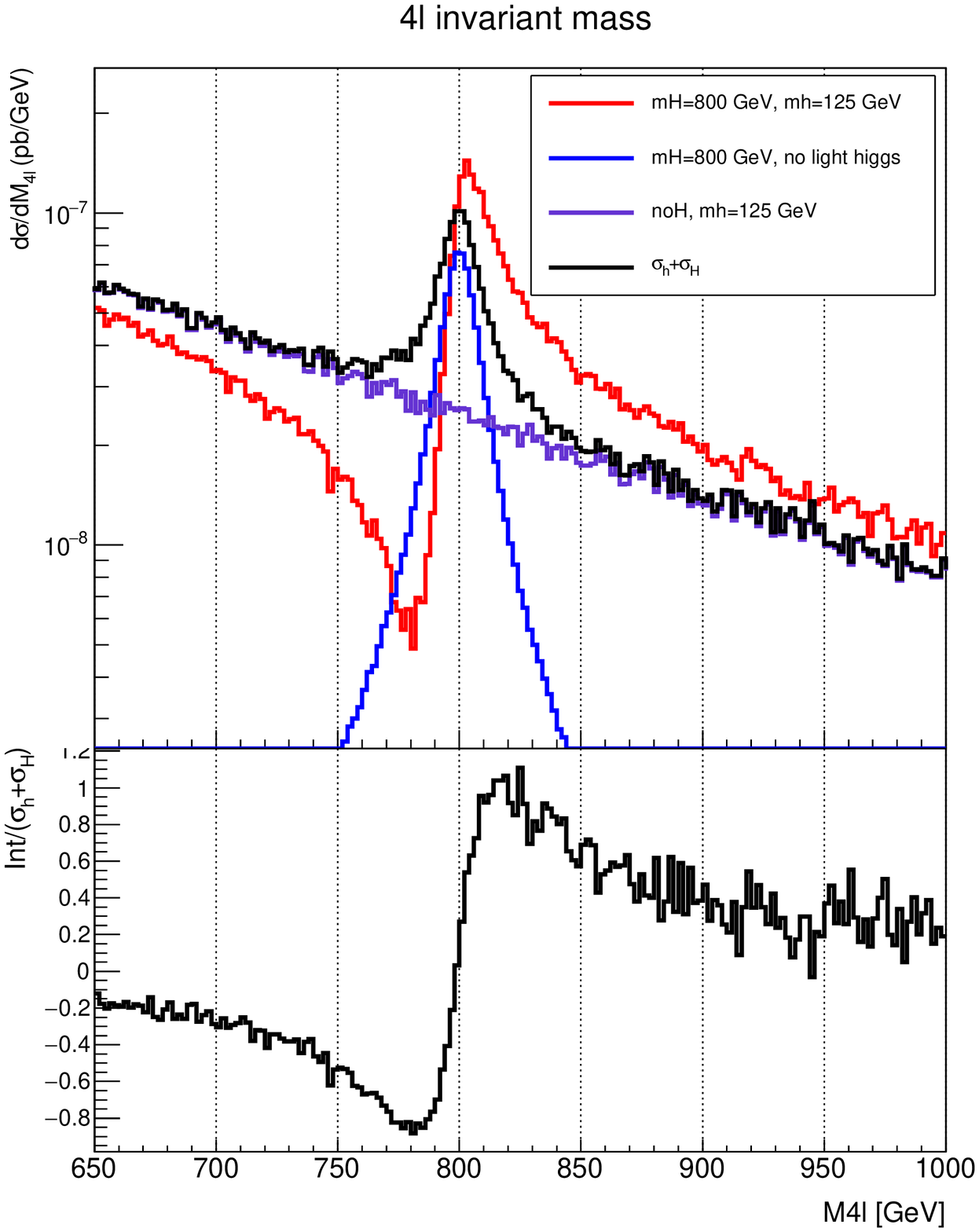}
\caption{Four lepton invariant mass distribution for $M_h=125$ GeV, $M_H=800$ GeV, $s_\alpha= 0.2$.
The meaning of the various histograms is as in \fig{fig:4lmass_400_02}.
}
\label{fig:4lmass_800_02}
\end{figure}

Our main results are shown in \figs{fig:4lmass_400_02}{fig:4lmass_800_02} and \tbn{table:sigma}. 
No cut has been applied to the final state. We have used CTEQ6L1 parton distribution functions
\cite{Pumplin:2002vw}. 
The ratio of vacuum expectation values $\tan\beta$ has been taken equal to one.

The width of the light and heavy scalar are 
\be
\Gamma_h = \Gamma^{SM}(M_h) c_\alpha^2, \qquad 
\Gamma_H = \Gamma^{SM}(M_H) s_\alpha^2 + \Gamma(H\rightarrow hh).
\ee
This corresponds, using \eqn{eq:H2hhWidth} for the $H\rightarrow hh$ width and 
\rf{Heinemeyer:2013tqa} for the SM Higgs width,
to $\Gamma_H = 1.77 (4.08)$ GeV
for $M_H = 400$ GeV, $s_{\alpha} = 0.2(0.3)$; $\Gamma_H = 15.80$ GeV
for $M_H = 600$ GeV and $s_{\alpha} = 0.3$;
$\Gamma_H = 16.69$ GeV for $M_H = 800$ GeV and $s_{\alpha} = 0.2$.
We have assumed $M_h = 125$ GeV which corresponds to $\Gamma^{SM}(125) = 4.03$ MeV.

\begin{table}[thb]
%\vspace{0.15in}
\begin{center}
\hspace*{-5mm}
\begin{tabular}{|c|c|c|c|c|c|c|}
\hline
& \multicolumn{3}{|c|}{200 GeV $< M_{4l} <$ 1 TeV} & 
\multicolumn{3}{|c|}{$M_H$$-$25 GeV $< M_{4l} <$ $M_H$+25 GeV}  \\
\hline
        & \hspace*{5mm} $\sigma$ \hspace*{5mm} 
        &  \hspace*{1mm} $\sigma_h + \sigma_H$ \hspace*{1mm}  
        & \hspace*{4mm} $\sigma_ H$ \hspace*{4mm}  
        & \hspace*{5mm} $\sigma$ \hspace*{5mm}  
        & \hspace*{1mm}  $\sigma_h + \sigma_H$ \hspace*{1mm}    
        &  \hspace*{3mm}  $\sigma_ H$   \hspace*{3mm} \\
\hline
$M_H$=400 GeV, $s_\alpha$=0.2 & 72.95 & 70.96 &  26.00  & 32.09  & 32.13  &  25.55 \\
\hline
$M_H$=400 GeV, $s_\alpha$=0.3  & 101.48  & 96.51 & 55.99  & 60.36  & 59.70  & 53.74 \\
\hline
$M_H$=600 GeV, $s_\alpha$=0.3  &  48.44 & 52.52  & 11.99   & 11.51  & 11.96  &  9.97\\
\hline
$M_H$=800 GeV, $s_\alpha$=0.2  & 43.96  & 45.96 & 1.00 &  1.57 &  1.46  &  0.65 \\
\hline
\end{tabular}
\end{center}
\caption{Cross sections in ab for $gg \rightarrow h,H \rightarrow ZZ \rightarrow 4e$
at the LHC with a center of mass energy of 13 TeV.}
\label{table:sigma}
\end{table}

We show the region around the heavy Higgs peak where the interference affects are most prominent.
The invariant mass distribution in the neighborhood of the light Higgs resonance is unaffected within 
the accuracy of the present simulation. 
In all figures the violet histogram is the result obtained including only the light Higgs with SM couplings 
scaled by $c_\alpha$ which we denote as $\sigma_h$. Since $\Gamma_h$ is small 
$\sigma_h\approx\sigma_{SM}(M_h) c^4_\alpha$ for large $M_{4l}$.
The blue line shows the cross section when only
the heavy Higgs is present, which will be referred to as  $\sigma_H$. 
The $H\rightarrow h h$ contribution to $\Gamma_H$ is retained and therefore  $\sigma_H$
is not equal to $\sigma_{SM}(M_H) s^4_\alpha$.
The full result, $\sigma$, is shown in red in \fig{fig:4lmass_400_02} for 
$s_\alpha = 0.2$, $M_H=400$ GeV,
in \figsc{fig:4lmass_400_03}{fig:4lmass_600_03} for $s_\alpha = 0.3$ with $M_H=400$ GeV
and $M_H=600$ GeV respectively and finally in \fig{fig:4lmass_800_02} for
$s_\alpha = 0.2$, $M_H=800$ GeV. 
The black histogram displays the sum of the blue and violet lines, $\sigma_{NI} = \sigma_h+\sigma_H$, 
and corresponds to neglecting the interference between the two scalars. The fractional size of the
correction to $\sigma_{NI}$ is
displayed in the bottom part of the figures where the ratio of the interference term and  $\sigma_{NI}$,
$\frac{\sigma - \sigma_{NI}}{\sigma_{NI}}$, is shown.

Contrary to naive expectations the light Higgs contribution is non negligible outside the peak region
\cite{Kauer:2012hd} and the interference effect is substantial. While the details depend obviously
on the mass of the heavy Higgs and on the mixing angle, we find a decrease of 10 to 20\% of
the differential cross section at invariant
masses of the four leptons of 300 GeV for $M_H=400$ and of about 20\% for $M_{4l}=350$,
$M_H=600$ and for $M_{4l}=500$, $M_H=800$.
This depletion becomes more pronounced as  $M_{4l}$ increases 
and reaches a dip which is almost
two orders of magnitude smaller than the predictions which neglect the interplay of the two Higgs fields.
The interference stays negative for four lepton masses below the heavy Higgs mass and then
turns positive.
It attains a maximun in which the true value is about a factor of two larger than $\sigma_{NI}$ and then
slowly decreases.  The position of the peak is shifted to slightly larger masses.
At four lepton masses about 200 GeV larger than $M_H$ the interference still amounts
to about 20 to 40\% of $\sigma_{NI}$. Because of unitarity, for very large $M_{4l}$ the full cross section
$\sigma$ must approach the SM result $\sigma_{SM}$. In this region, where the width of the two
Higgses can be neglected, $\sigma_{SM}=\sigma_h/ c^4_\alpha=\sigma_H/ s^4_\alpha$.

\tbn{table:sigma} shows the cross section in ab for two mass intervals:  
200 GeV $< M_{4l} <$ 1 TeV, which roughly coincides with the range employed so far by the experimental
collaborations to set limits on the presence and couplings of additional scalars,
and $M_H$$-$25 GeV $< M_{4l} <$ $M_H$+25 GeV, as an indication of the possible effects on an
analysis in smaller mass bins which requires high luminosity.
In the first case, the contribution of the heavy Higgs is a relatively small fraction of the Higgs production
cross section in gluon fusion.
Furthermore, the interference effects in the 200 GeV $< M_{4l} <$ 1 TeV depend crucially on the heavy
Higgs mass. For $M_H=400$ the exact result is larger than the incoherent sum
$\sigma_h + \sigma_H$. The long tail for  $M_{4l}  > M_H$ gives a larger contribution than the
intermediate region 200 GeV $< M_{4l} < M_H$. On the contrary, for larger heavy Higgs masses we have
$\sigma < \sigma_h + \sigma_H$. The negative interference in the intermediate region outweights
the positive contribution at larger masses.
In the smaller range,  $M_H$$-$25 GeV $< M_{4l} <$ $M_H$+25 GeV, the positive and negative contributions are very close for $M_H=400$ GeV and the full result is in rough agreement with 
$\sigma_h + \sigma_H$. For this value of $M_H$ and mass interval, the heavy Higgs contribution 
is significantly larger than the light Higgs one. 
For $s_\alpha = 0.3$, $M_H=600$ GeV the exact cross section is about 4\% smaller than the incoherent sum while for $s_\alpha = 0.2$, $M_H=800$ GeV it is approximately 8\% larger.
We see that for large values of the heavy Higgs mass the interference effects are non negligible even on a restricted mass range.

Our results have no pretense to be a complete prediction. They need to be validated with the inclusion of
the continuum contribution and of higher order corrections.
It should be noticed that, for Higgs decay to color neutral final states, all relevant amplitudes in
QCD will have the structure of \eqn{eq:higgsExchange} and therefore interference effects between the
scalar fields will not be spoiled by QCD corrections. 

The interference with the quark box amplitude
deserves more care. The gluon-gluon continuum term does not involve scalar exchanges and therefore
it cannot be cast in the form of
\eqn{eq:higgsExchange} and could in principle dilute the effect. However, the interference
between the box diagrams and the light Higgs mediated ones is always negative,
while the amplitude with a
heavy Higgs exchange changes sign at the resonance. As a consequence the continuum
and the heavy Higgs term are expected to be in phase for $M_h < M_{4l} < M_H$ and out of phase
for $M_{4l} > M_H$.
A detailed study of this topic is in preparation.
  
\section{Conclusions}
In any theory with  multiple neutral Higgs which couple to the same set of elementary particles 
the scalars  are expected to interfere. 
We have shown in the case of Higgs production through gluon fusion
in the 1SHM that the interference effects can be significant and cannot be neglected when aiming for high
accuracy  predictions.
 
\section *{Acknowledgments}

Daily discussions with Alessandro Ballestrero have been invaluable and are gratefully acknowledged. \\
This work has been supported by MIUR (Italy) under contract 2010YJ2NYW\_006  and by the Compagnia di
San Paolo under contract  ORTO11TPXK.\\

%\hfill*
%\eject
%\newpage

%%%%%%%%%%%%%%%%%%%%%%%%%%%%%%%%%%%%%%%%%%%%%%%%%%%%%%%%%%%%%%%%%%%%%%%%%

% To include bibliography do:
% 1- pdflatex FileName.tex
% 2- bibtex FileName
% 3- pdflatex FileName.tex
% 4- pdflatex FileName.tex
%\bibliographystyle{unsrt}
%\bibliographystyle{hunsrt}
%\bibliographystyle{h-elsevier.bst}

\bibliographystyle{JHEP}

\bibliography{MySinglet}

\end{document}